\documentclass[twocolumn,prl,superscriptaddress,showpacs]{revtex4}
\usepackage{epsfig,rotating}
\usepackage{multirow,bm,amsmath,amssymb}
\usepackage{graphicx}
\usepackage{mathrsfs,color}

\newcommand{\beq}{\begin{equation}}
\newcommand{\eeq}{\end{equation}}
\newcommand{\beqn}{\begin{eqnarray}}
\newcommand{\eeqn}{\end{eqnarray}}

\newcommand{\bsub}{  \begin{subequations}}
\newcommand{\esub}{ \end{subequations}}

\newcommand{\blue}[1]{\textcolor{black}{#1}}

\newcommand{\red}[1]{\textcolor{black}{#1}}

\begin{document}

\title{Anharmonicity of multi-octupole-phonon excitations in $^{208}$Pb:
analysis with multi-reference covariant density functional theory and subbarrier fusion of $^{16}$O+$^{208}$Pb}
\author{J. M. Yao}
   \affiliation{School of Physical Science and Technology, Southwest University, 400715 Chongqing, China}
  \affiliation{Department of Physics, Tohoku University, Sendai 980-8578, Japan}
 \affiliation{Department of Physics and Astronomy, University of North Carolina, Chapel Hill, NC 27516-3255, USA}

\author{K. Hagino}
  \affiliation{Department of Physics, Tohoku University, Sendai 980-8578, Japan}
\affiliation{Research Center for Electron Photon Science, Tohoku
University, 1-2-1 Mikamine, Sendai 982-0826, Japan}
\affiliation{ National Astronomical Observatory of Japan, 2-21-1 Osawa, Mitaka, Tokyo 181-8588, Japan}


\begin{abstract}
We discuss anharmonicity of the multi-octupole-phonon states
in $^{208}$Pb based on a covariant density functional theory, by fully taking into account the interplay
between the quadrupole and the octupole degrees of freedom. Our results indicate the existence of
a large anharmonicity in the transition strengths, even though the excitation energies are similar to those in the harmonic
limit. We also show that the quadrupole-shape fluctuation significantly enhances the fragmentation of the two-octupole-phonon states in $^{208}$Pb. Using those transition strengths as inputs to coupled-channels calculations, we then discuss
the fusion reaction of $^{16}$O+$^{208}$Pb at energies around the Coulomb barrier. We show that the anharmonicity
of the octupole vibrational excitation considerably improves previous coupled-channels calculations in the harmonic
oscillator limit, significantly reducing the height of the main peak in the fusion barrier distribution.
\end{abstract}


\maketitle

Collective vibrational excitations exist commonly in many-fermion systems \cite{BB94}.
Here, the concept of {\em phonon} is an important keystone
to understand these excitations. For instance, for finite nuclear systems,
vibrations of the nuclear surface are treated as elementary excitations \cite{BM75,Ring80}.
The phonons for these vibrations are boson-like in character and multiple
excitations of the same type are possible, resulting in {\em multi-phonon} states~\cite{BM75}.
In a harmonic vibration, all the levels in a phonon multiplet are
degenerate in energy and the energy spacing between neighboring multiplets
is a constant. The energy patterns close to such harmonic vibration
have been observed in nearly spherical nuclei in different mass regions, and these states have been primarily interpreted
as multipole phonon states~\cite{BM75,Aumann98,Corminboeuf00}.

From a microscopic viewpoint, however, the collective excitations are generated by a coherent superposition of quasiparticle excitations of fermions in orbits close to the Fermi surface. The Pauli principle correction, together with residual interactions among phonons (that is, mode-mode couplings), modifies the structure of multi-phonon states and make them
highly fragmented \cite{Brink65}. To \blue{assess} the degree of such anharmonicity
is a fundamental question in nuclear physics, that needs to be studied
more extensively. In particular, we mention that recent studies on the anharmonicity in the multi-quadrupole-phonon states have questioned the concept of low-energy vibrational modes in atomic nuclei~\blue{\cite{Garrett08,Perez15}}.

The double-magic nucleus $^{208}$Pb provides an ideal laboratory to examine the concept of {\em multi-octupole-phonon}
excitations in nuclear systems, as the first excited $3^-$ state of this nucleus has long been interpreted as a collective
one-octupole-phonon state~\cite{Hamamoto}.
In the past decades, several experimental searches for the {\em two-octupole-phonon} (TOP)
states in $^{208}$Pb have been carried out~\cite{Yeh96,Yeh1998,Vetter98,Valnion01}.
Even though many of the TOP members have
been identified, the multi-phonon excitations in $^{208}$Pb have not yet been understood completely.
This is the case especially for the nature of the TOP multiplets, which has been predicted
to show a strong fragmentation~\cite{Valnion01,Ponomarev99,Brown00}.

Incidentally, heavy-ion fusion reactions provide an alternative way to probe the multi-phonon excitations
in atomic nuclei, which significantly affect the subbarrier fusion cross sections~\cite{HT12,DHRS98,BT98,Back14}
as illustrated by coupled-channels calculations~\cite{HT12,Hagino97,Hagino98}.
Previous coupled-channels calculations for the fusion reaction of $^{16}$O+$^{208}$Pb based on
multi-harmonic-phonon excitations fail to reproduce the observed energy dependence of fusion cross
sections~\cite{Morton99,Esbensen07,YHR12}, and overestimate the height of the main peak in
the so called fusion barrier distribution~\cite{RSS91}. It has been a long-standing unsolved question
how the fusion cross sections for the $^{16}$O+$^{208}$Pb system can be accounted for by
the coupled-channels approach.

In this paper, we for the first time examine the concept of multi-octupole-phonon excitations in $^{208}$Pb
in the microscopic framework of generator coordinate method (GCM) based on a covariant energy
density functional~\cite{Niksic11}. This beyond mean-field approach is also referred to as multi-reference
covariant density functional theory and has been rapidly developed in the past decade~\cite{Yao10,Yao15-Ra,Zhou15}.
In this method, collective vibrational excitations are described as fluctuations in nuclear shapes in a full microscopic manner
and therefore the Pauli principle correction to the
phonon excitations is taken into account automatically.
We show that this method is capable to capture the main characters of
nuclear multipole phonon excitations, which are generally fragmented by
its internal fermionic structure and by coupling to
other shape degrees of freedom. We show that these anharmonic features
considerably improve the coupled-channels calculations for the fusion reaction of $^{16}$O+$^{208}$Pb,
significantly reducing the height of the main peak in the fusion barrier distribution.

In the multi-reference covariant density functional theory,
the wave functions for nuclear collective states are constructed
by superposing a set of quantum-number projected nonorthogonal mean-field reference states
$\vert \beta_2\beta_3\rangle$ around the equilibrium shape. Here,
the reference states $\vert \beta_2\beta_3\rangle$
are obtained by deformation constrained relativistic mean-field calculations with
the quadrupole and octupole deformation parameters, $\beta_2$ and $\beta_3$,
respectively. The wave functions thus read
\begin{equation}\label{gcmwf}
\vert JM\pi\rangle
=\sum_{q} f^{J\pi}(q)
\hat P^J_{M0} \hat P^N\hat P^Z  \hat P^\pi\vert q\rangle,
\end{equation}
where $q$ refers to $(\beta_2,\beta_3)$ and $\hat P$'s
are projection operators onto the angular momentum $J$, the
parity ($\pi=\pm$), and the neutron and proton numbers ($N, Z$)~\cite{Ring80}.
For the sake of simplicity, we have restricted all the
reference states $\vert q\rangle$ to be axially deformed.
\blue{We note that the effects of pairing vibrations are
not taken into account in the present calculation, even though they
may have an influence on low-lying excited $0^+$
states~\cite{Bortignon77}.} The weight function $f^{J\pi}(q)$ in Eq. (\ref{gcmwf})
as well as the energy $E^\pi_{J}$ for each GCM state
$\vert JM\pi\rangle$ are determined by the Hill-Wheeler-Griffin equation~\cite{Hill57}.
As in our previous studies~\cite{Yao10}, the mixed-density prescription
is adopted for the energy overlap in the Hamiltonian kernel.
In the calculations presented below, we employ the relativistic energy functional PC-F1~\cite{PC-F1}.
The pairing correlation among the nucleons is treated in the BCS approximation with a density-independent
$\delta$-force supplemented with a smooth energy
cutoff~\cite{Krieger90}. The strength parameters in the pairing force
are chosen according to the PC-F1 force~\cite{PC-F1}.

\begin{figure}[bt]
\includegraphics[width=7cm]{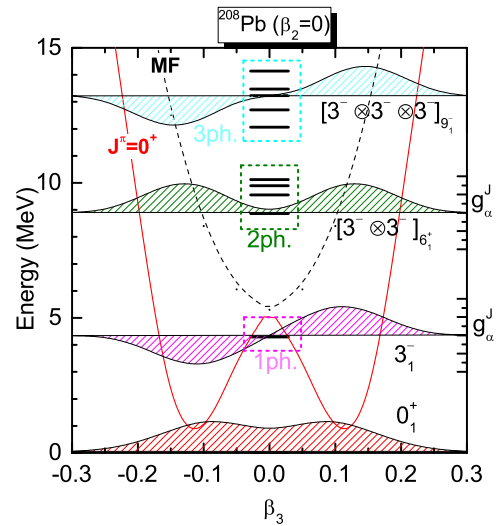}
\caption{(Color online) The total energy (normalized to the GCM ground state)
of the $^{208}$Pb nucleus as a function of octupole deformation parameter $\beta_3$.
The quadrupole degree of freedom is frozen at $\beta_2=0$. The energy curve with projection onto the
particle numbers ($N, Z$) and the spin-parity of $0^+$
is also shown by the solid line, together with the excitation energies and the collective wave functions
for the states. } 
\label{MOP}
\end{figure}

We first examine the concept of one-dimensional vibration in $^{208}$Pb with octupole degree of freedom,
by freezing the quadrupole degree of freedom at $\beta_2=0$. The deformation energy curve in the mean-field approximation as
well as the projected energy curve for $J^\pi=0^+$ are shown in Fig.~\ref{MOP}. The excitation energies of natural-parity ``phonon" states and the collective wave functions defined by $g^{J\pi} (q) \equiv \sum_{q'}
\left[\mathscr{N}^{J\pi}_{q, q'}\right]^{1/2}f^{J\pi}(q')$
for the $0_1^+$, $3_1^-$, $[3^-\otimes3^-]_{6^+}$,
and $[3^-\otimes3^-\otimes3^-]_{9^-}$ states are also shown, where
$\mathscr{N}^{J\pi}_{q, q'}$ is the norm kernel in the
Hill-Wheeler-Griffin equation~\cite{Yao15-Ra}.
The mean-field energy curve is almost parabolic and
is centered at $\beta_3=0$. As expected, the dynamical octupole
effect \red{originated} from the symmetry restoration generates
two octupole minima and shifts the dominant component in the $0^+$ state
to an octupole-deformed configuration with $\vert\beta_3\vert\sim 0.1$.
One can also see that the multiples of two- and three-``phonon" states appear
at similar excitation energies to each other.
The spin-average of the excitation energies for the natural-parity two-
and three-octupole phonon states is 9.6 MeV and 13.1 MeV, respectively, which is about twice and three times
the energy of the one-octupole phonon state, 4.3 MeV, \red{close to $\sim4.0$ MeV from non-relativistic GCM
calculations~\cite{Heenen01,Robledo14} (the excitation
energies are expected to be lowered down if the cranking mean-field states are adopted
in the configuration mixing calculations~\cite{Li12,Borrajo16})}.
Notice, however, that there are some energy displacements, indicating the existence
of anharmonicity. The energy displacement appears to increase with the number
of the ``phonons" in the states. As can be seen in the figure, the wave
functions for the $0^+_1$ and $3^-_1$ states show a similarity to the wave
functions for the zero and one-phonon states of a harmonic oscillator, while
those of the $[3^-\otimes3^-]_{6^+}$
and $[3^-\otimes3^-\otimes3^-]_{9^-}$ states are considerably distorted.

\begin{figure}[bt]
\begin{center}
\includegraphics[width=9cm]{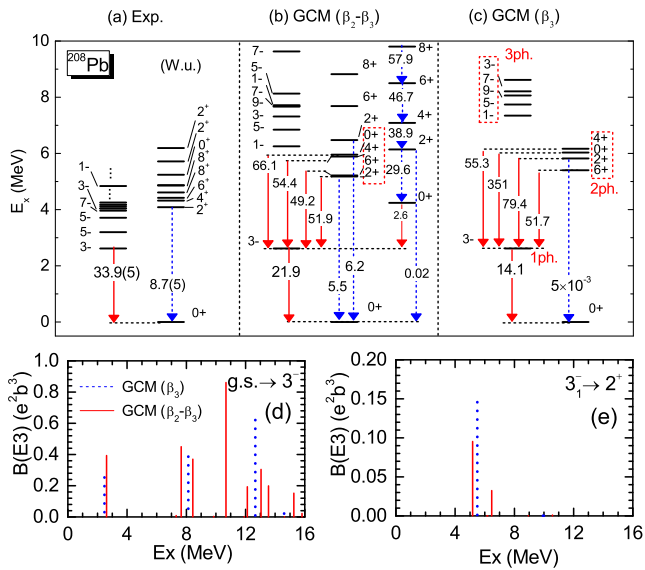}
\caption{(Color online) The low-lying energy spectra of $^{208}$Pb
obtained by mixing the octupole-quadrupole ($\beta_3-\beta_2$)
deformed configurations (the panel (b)) and by mixing only
the octupole ($\beta_3$)
deformed configurations (the panel (c)), in comparison with the
experimental data taken from Ref. \cite{NNDC} (the panel (a)).
The panels (d) and (e) show the  $E3$ transition strengths from the
the ground state to $3^-$ states and from the first $3^-$ state
to $2^+$ states, respectively, as a function
of the excitation energy of the final states.
In the panels (a)-(c), the red solid and the blue dashed lines
indicate the $E3$ and $E2$
transition strengths (in W.u.), respectively.
All the calculated excitation energies are scaled to the empirical excitation
energy of the lowest $3^-$ state by dividing them by a constant factor. }
\label{spectrum}
\end{center}
\end{figure}

It is interesting to notice that an anharmonicity is stronger in
the transition strengths as shown in Fig.~\ref{spectrum}(c),
where the $E3$ transition strength from the $3^-_1$ state to the
ground state is underestimated by a factor of more than two.
Moreover, the $E2$ transition strength from the first $2^+$ state
to the ground state is underestimated by three orders of magnitude.
The strength of the $E3$ transition from the two-phonon multiplets to
the $3^-_1$ state is also much larger than twice the $B(E3)$ value
from the $3^-_1$ state to the ground state. In particular,
the $E3$ transition from the $[3^-_1\otimes3^-_1]_{0^+}$ state
to the 3$^-_1$ state is much stronger than that from the other
multiplets  of the TOP states. We note that a large anharmonicity in the transition strengths
has been found also in the ``multi-quadrupole-phonon" excitations
in \blue{Refs.~\cite{Garrett08,Perez15}}.

To examine the anharmonicity arising from the coupling between the octupole and the quadrupole
shape fluctuations, we next carry out the GCM calculation in the two-dimensional $(\beta_2, \beta_3)$ deformation plane.
The calculated low-lying energy spectra are shown in Fig.~\ref{spectrum}(b), where only natural-parity states are plotted.
Here, the spectra are scaled by a constant factor
so that the energy of the first 3$^-$ state matches with
the empirical value, 2.62 MeV. Notice that the inclusion of the quadrupole shape fluctuation
slightly alters the excitation energies of the $[3^-_1\otimes3^-_1]_{0^+, 2^+, 4^+, 6^+}$
states. One can see that, after including this effect,
the transition strengths for the $3_1^- \to 0^+_1$ and
the $2_1^+ \to 0^+_1$ transitions are \blue{closer to the experimental
data. The $B(E3; 3^-_1 \to 0^+_1)$ value is consistent with
 21.93 W.u. by the Gogny D1S force~\cite{Robledo14},
although it is slightly larger than 18.7 W.u by the
Skyrme SLy4 force~\cite{Heenen01}. Moreover,}
 the electric dipole transition strengths are also
well reproduced by this calculation. For instance, we obtain
the $B(E1)$ value from the first 4$^+$ state to the first 5$^-$ state
to be 1.5$\times 10^{-4}$ W.u., which is compared to the experimental
upper bound of 1.0$\times 10^{-4}$ W.u. \cite{Yeh1998}.
We note that, in the octupole-quadrupole shape-mixing calculation,
the $E3$ transition strengths from the four TOP states to
the $3^-_1$ state are similar to each other, about three times
the $B(E3)$ value from the $3^-_1$ state to the ground state.
Figures~\ref{spectrum}(d) and (e) show the calculated $B(E3)$
values from the ground state to 3$^-$ state and those from  the first 3$^-$ state to $2^+$ states as
a function of the excitation energy of the final states, respectively.
The solid and the dotted lines indicate the results of
the octupole-quadrupole shape fluctuation and of the octupole vibration
only, respectively. One can see that generally the $E3$ transitions
become more fragmented after taking into account the
fluctuation in the quadrupole shape. In particular, the $E3$ transition
from the $3^-_1$ state to excited $2^+$ states are
strongly quenched. Similar phenomenon is also found in the $E3$ transitions
from the $3^-_1$ state to other excited states (not shown here).

\begin{figure}[]
\begin{center}
\includegraphics[width=8.5cm]{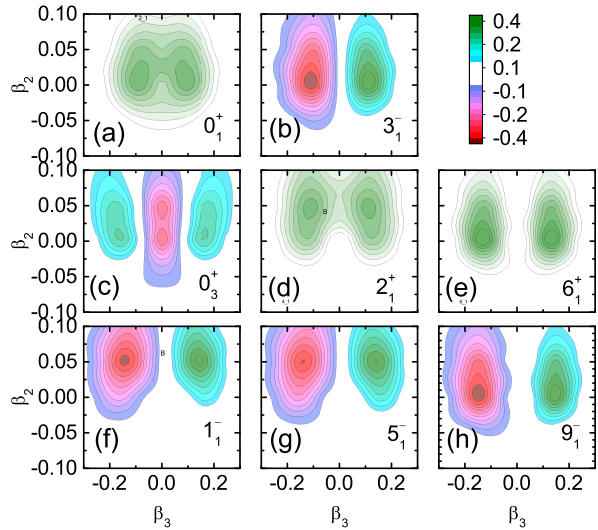}\vspace{-0.5cm}
\caption{(Color online) The distribution of the collective wave
functions $g^{J\pi}$ for several selected low-lying states
of $^{208}$Pb shown in Fig.~\ref{spectrum}(b).}
 \label{collwf2}
\end{center}
\end{figure}

Figure~\ref{collwf2} shows the distribution of the collective wave
functions for the ground state and some selected excitation
states. Comparing with the wave functions in Fig.~\ref{MOP},
one can see a rather large fluctuation along the quadrupole
deformation in all the states. For the multiplets of the TOP states,
only the $0^+$ state has two nodes along $\beta_3$ direction, showing again the anharmonicity
in the wave functions.

\begin{figure}[tb]
\includegraphics[width=7.5cm]{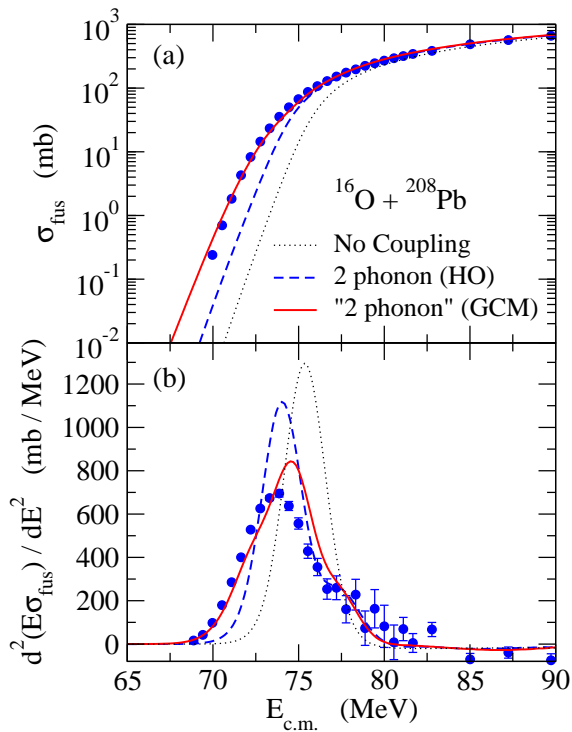}
\caption{(Color online) The fusion cross sections (upper panel) and the fusion
  barrier distributions (lower panel) for the $^{16}$O+$^{208}$Pb
  system obtained with the semi-microscopic coupled-channels calculation
with the coupling strengths from the MR-DFT calculations (the solid line).
The dashed and the dotted lines show the results of the two-phonon coupling in
the harmonic oscillator limit and of the no-coupling limit, respectively.
The experimental data are taken from Ref.~\cite{Morton99}.
}
\label{fig:fusion}
\end{figure}

\blue{We have repeated the GCM calculation using the PC-PK1
force \cite{PC-PK1}.
The calculated excitation energies and electric multipole
transitionstrengths for the one- and two-octupolephonon states
turned out to be similar to those by the PC-F1 force.
However, we have found that the $2^+$ states and high-lying
states with the PC-PK1 force may have a problem of convergence, as they
are much more sensitive to
the model space than those by the PC-F1 force,
even though
both PC-F1 and PC-PK1 forces give much better convergent solutions
for the quadrupole-phonon states in $^{58, 60}$Ni \cite{Hagino15}.
In view of this, we present
only the PC-F1 results in this paper. }

In order to further test the results of the present calculation
for the $^{208}$Pb nucleus, we next discuss the sub-barrier fusion reaction
of the $^{16}$O+$^{208}$Pb system. To this end, we employ the semi-microscopic
approach \cite{Hagino15} and solve the coupled-channels equations by using
the transition strengths from the GCM calculations as inputs.
In this approach, the internuclear potential and the coupling potentials
are generated from a phenomenological deformed Woods-Saxon potential.
For this, we use the parameters of $V_0$=178 MeV,
$R_0$=0.978 $\times(16^{1/3}+208^{1/3})$ fm, and $a$=1.005 fm, which are
similar to those used in Ref. \cite{Morton99}.
In the coupled-channels calculations, in addition to the entrance channel,
we include the one-octupole phonon state, 3$^-_1$, at 2.615 MeV,
the ``one-quadrupole'' phonon state, 2$^+_1$, and several states which are
strongly coupled to those 3$^-_1$ and 2$^+_1$ states by the octupole and
the quadrupole couplings. The whole TOP candidate
states are included in this model space.
As is shown in Ref. \cite{Hagino15}, we scale all the excitation energies to
the empirical excitation energy of the 3$^-_1$ state. We also scale
all the coupling strengths to the empirical coupling strength between the ground state and
the 3$^-_1$ state, that is, $\beta=0.144$, which is estimated from the
measured $B(E3)$ strength with the radius parameter of $r_0$=1.1 fm.
The resultant coupled-channels equations are solved using the computer
code {\tt CCFULL} \cite{HRK99}.

The solid line in Fig. \ref{fig:fusion} is the fusion cross sections (the
upper panel) and the fusion barrier distribution (the lower panel) so obtained.
Here, the fusion barrier distribution is defined as the second energy
derivative of the product of the energy $E$ and fusion cross section $\sigma_{\rm fus}$,
that is, $d^2(E\sigma_{\rm fus})/dE^2$ \cite{DHRS98,RSS91}. This is compared to the two-phonon
calculations in the harmonic oscillator
limit (the dashed line) and to the single-channel calculation (the dotted
line). For the former, we include the 3$_1^-$, 2$_1^+$,
$(3_1^-)^2$, $(2_1^+)^2$, and $3_1^-\otimes 2_1^+$ states within the harmonic
oscillator coupling scheme \cite{HT12}.
It has been a long-standing problem that for this particular system
the coupled-channels calculation with the harmonic oscillator couplings
overestimates the height of the main peak in the barrier
distribution \cite{Morton99,Esbensen07,YHR12}.
It is remarkable that the present GCM calculation yields a much lower
peak in the fusion barrier distribution, leading to a much better agreement with the experimental data both for the fusion
cross sections and for the barrier distribution.
For this good reproduction, it turns out that the coupling between the 3$^-_1$ and
the 2$^+_1$ states, as well as the couplings between the TOP states and the excited negative parity states,
play an important role. In the previous coupled-channels calculations,
the 3$^-$, 2$^+_1$ and the 5$^-_1$ states have been treated as
independent phonon states, and these couplings were absent in the calculations. In contrast,
in the present GCM calculation, the 2$_1^+$ and the $5_1^-$ states
have in part the two and three octupole phonon characters, respectively.
Likewise, the 1$^-$ states have both
the $(3_1^-)^3$ and the $3_1^-\otimes 2_1^+$ characters. Apparently those anharmonicity
effects in the transition strengths
lead to the strong couplings between the ground state and those
states via multiple octupole excitations, significantly improving the
previous coupled-channels calculations.

In summary, the multi-octupole-phonon excitations in $^{208}$Pb
have been examined with the multi-dimensional GCM calculations based on a covariant energy density functional.
We have shown that the coupling to quadrupole shape fluctuation leads to
a stronger fragmentation of the double-octupole phonon states and also enhances the $E3$ transition
strength between the ground state and the single-octupole
phonon state. These calculated transition
strengths have then been used as inputs to the coupled-channels
equations in order to discuss the sub-barrier fusion
reaction of $^{16}$O+$^{208}$Pb. We have shown that these
anharmonicities in the transition strengths play an important role in this reaction, leading
to a much better reproduction of fusion barrier
distribution as compared to the previous coupled-channels calculations.

Our calculations indicate that anharmonicity of nuclear vibrations
is much larger in the collective wave functions and in the
transition properties as compared to the anharmonicity in the excitation
energies. An interesting feature is that the anharmonicity may be large even if the energy spectrum resembles
a harmonic oscillator. It will be interesting to reexamine systematically nuclear
vibrations with a fully microscopic theory such as the multi-reference
density functional approaches \blue{based on both nonrelativistic and relativistic energy functionals in future.}

\bigskip

We thank D. J. Hinde and T. Ichikawa for useful discussions. This work was partially supported by the NSFC under Grant Nos. 11575148, 11475140, and 11305134, and JSPS KAKENHI Grant Number 2640263.


\end{document}